# Hydrodynamic and Solid Mechanics Analysis of Capillary Force-Induced Mold Deformation in Sub-10 nm UV Nanoimprint Lithography

**Jingxuan Cai[1] and Wen-Di Li[1]**

[1] Department of Mechanical Engineering, The University of Hong Kong, Pokfulam, Hong Kong, China

E-mail: liwd@hku.hk

**Abstract**

A model has been developed to study the dynamic filling process and to investigate the capillary force-induced deformation of nanostructures on the imprint mold during ultraviolet nanoimprint lithography (UV-NIL) down to sub-10 nm resolution. The dynamic behavior of resist filling with varied physical parameters was investigated by a hydrodynamic model. The capillary force-induced deformation of mold structures was modeled using beam bending mechanics for both wetting and non-wetting mold structures. Theoretically calculated results were cross-validated with finite-element simulations using two-phase flow and solid mechanics methods. Based on the theoretical analysis, a general parameter of critical aspect ratio for design of imprint mold for UV-NIL is developed. The investigation of capillary force-induced deformation in UV-NIL helps to deepen the understanding of dynamic mechanism of resist filling and structural deformation at sub-10 nm scale and enable optimization for high-fidelity UV-NIL.

## 1. Introduction

Nanoimprint lithography (NIL) is a promising technology that enables cost-effective and high-throughput fabrication for micro- and nanopatterns at high patterning resolution. One of the most advanced NIL techniques under intensive study is the ultraviolet NIL (UV-NIL). During UV-NIL process, a liquid UV-curable resist is spin-coated or dispensed on a substrate in the atmospheric environment. Then the resist is covered by a mold with nanopatterns and the resist liquid will fill the nanocavities due to the capillary force. The resist is cured under ultraviolet exposure afterward, leaving solidified nanoimprinted patterns on the substrate. Fabrications of nanostructures with high transfer fidelity using the UV-NIL have been widely reported, and the patterning resolution could achieve at 10 nm when a helium ion beam lithography (HIBL) patterned hydrogen silsesquioxane (HSQ) mold was used, because of the excellent patterning resolution and high Young's modulus of focused helium ion beam exposed HSQ.[1-4] However, pattern deformation and structure collapse arose as essential issues when advancing the UV-NIL toward sub-10 nm resolution.[1] Therefore, general understanding of the resist filling process and the deformation of mold structures during the UV-NIL is highly demanded for the design and optimization of the imprint mold for the use in sub-10 nm resolution UV-NIL. Concurrently, numerous efforts have been made to investigate the filling process[5-7] and bubble formation and dissolution process[8-10] for the UV-NIL using hydromechanic analysis, as well as molecular dynamics simulation. [11] However, with the further shrinkage of structure dimensions, the structures of the imprint mold would suffer from deformations due to the dramatically increased capillary force[1] and thus limits the further application of UV-NIL. Although the resist deformation for NIL has been investigated using solid mechanics[12] and molecular dynamics methods, [13] and the mold deformation for thermal NIL has been studied, [14] to our best knowledge, the general model of the capillary force induced pattern deformation on the mold for UV-NIL towards sub-10 nm resolution has not been reported yet.

Herein, we have proposed a theoretical model, which incorporates both the hydrodynamics and solid mechanics, to analyze stress distribution in the mold structures during the UV-NIL process through modeling the dynamic filling process of the resist flow for both wetting and non-wetting mold materials and cross-validated the theoretical modeling with the results obtained from a numerical simulation using the finite-element method (FEM). Furthermore, the capillary force-induced deformation of mold structures in UV-NIL with resolution down to 10 nm was theoretically investigated. Based on the analysis, a general critical aspect ratio of imprint mold applicable was proposed and estimated for imprint molds with different yield strengths.

## 2. Model and simulation

The model discussed here has been applied to a system in which two parallel beams of width $w$ are separated by a space of gap $d$, the height of the beams is $H$. The models for the wetting and non-wetting imprinting process are schematically demonstrated in **Figure 1a** and **1b**, respectively. In a typical UV-NIL process with a wetting mold, when the mold approaches the substrate, the capillary force pulls the resist liquid up to the features fabricated on the template. The fluidic flow of the resist can be described as a Newtonian fluidic. For simplification, the mold is considered to be stationary, and the whole system is assumed to be in a standard atmospheric pressure. The elastic deformation of beam structures before collapse during UV-NIL is ignored to simplify the model.

The pressure difference across the liquid-air interface can be estimated by Young-Laplace equations:[15-17]

$$p_0 - p_i = \gamma(\frac{1}{r_1} + \frac{1}{r_2}) \quad (1)$$

where $\gamma$ is the surface tension of the resist, and $r_1$ and $r_2$ are the principal radii of curvature.

while

$$r_1 = \frac{d/2}{cos\theta} \quad (2)$$
$$r_2 = \infty \quad (3)$$

where $\theta$ is the contact angle of resist with substrate. Then

$$p_i = p_0 - \frac{2\gamma cos\theta}{d} \quad (4)$$

As the deformation of mold structures before collapse is considered to be neglected, the capillary rise process of resist liquid can be derived from a Poiseille flow by the Lucas-Washburn equation:[18-24]

$$h(t) = \left(\frac{\gamma d\, cos\,\theta}{3\mu}\right)^{1/2} \sqrt{t} \quad (5)$$

Beams of the mold are considered as elastic material for evaluation of the capillary force-induced deformation, and the yield strength is used as the collapse criterion. Capillary force induced stress is distributed inside the beams corresponding to the pressure difference across the beam. The model structure for calculating the capillary force induced stress distribution is schematically illustrated in **Figure 1c**. The momentum of inertia is

$$I = \frac{1}{12}Dw^3 \quad (6)$$

where $D$ is the length of the beam. The bending momentum at the cross section is

$$M(h) = \int_0^h -q(z)(H-z)D\, dz \quad (7)$$

Here, the load on the beam induced by the pressure difference across the beam is assumed to be uniform, then

$$q(z) = p_0 - p_i \quad (8)$$

From the elastic beam bending model, the maximum stress across the beam occurs at the corner of the beam, which is

$$\sigma_{max}(h) = \frac{M(h)}{I} \cdot \frac{w}{2} = \frac{6\gamma cos\theta}{d} \cdot \frac{2Hh - h^2}{w^2} \quad (9)$$

The maximal value occurs at $h = H$, which could be estimated by,

$$\sigma_{max} = \frac{6\gamma\, cos\,\theta}{d}\left(\frac{H}{w}\right)^2 \quad (10)$$

In case of UV-NIL using a non-wetting mold, the surface tension prevents resist liquid from filling the cavities. Therefore, external pressure is required to overcome the surface tension and push the resist liquid into channels of the mold structures. The model for the non-wetting imprint process is schematically illustrated in **Figure 1b**. The pressure difference across the liquid-air interface, according to the Young-Laplace equations for the forward contact angle $\theta$, which is the critical pressure maintaining the hydrostatic force balance, is given by

$$p_r - p_0 = \frac{2\gamma\, cos(\pi - \theta)}{d} \quad (11)$$

When the external pressure exceeds the critical pressure, the resist liquid will fill the channels due to the pressure difference.

Similar to the wetting-wall model, the maximum stress across the beam could be estimated by,

$$\sigma_{max} = \frac{6\gamma\, cos(\pi - \theta)}{d}\left(\frac{H}{w}\right)^2 \quad (12)$$

Finite element method (FEM) simulations are performed using the level-set method and coupled with solid mechanics to validate theoretical modeling. The modeling structure is schematically illustrated in **Figure 1d**. The integration time step is 0.025 μs and the total simulation time is usually within a few microseconds. The contact angles between the walls and the resist are 30°, and the upper and lower boundaries are fixed. Left and right boundaries are the inlets, and the deformation of the structures is neglected in the simulation process using fluidic dynamics. The simulation of solid mechanics is then implemented to simulate the mechanical behavior of mold structures by applying the pressure difference obtained in the previous simulation of fluidic dynamics. The simulations of fluidic dynamics and solid mechanics are conducted using the front-tracking two-phase flow (TPF) method and solid mechanics modules of COMSOL Multiphysics software, respectively.

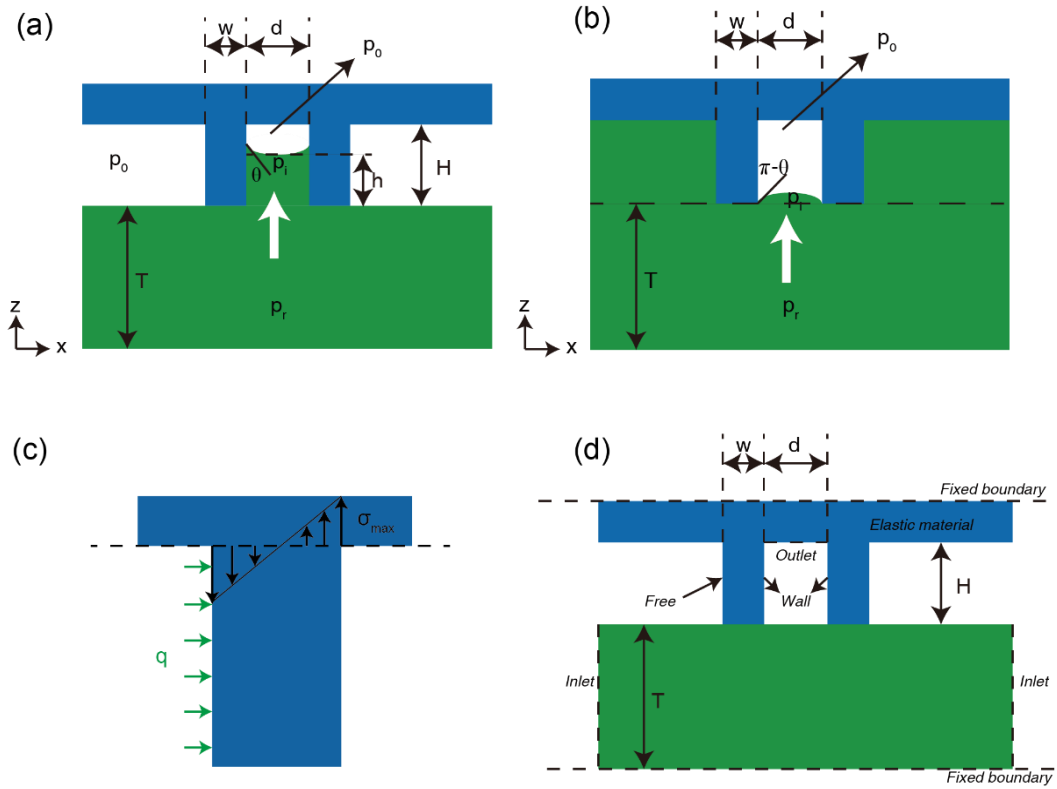

Figure 1 Schematic of the model for the (a) wetting and (b) non-wetting molds. *T* is the thickness of the resist, *H* and *w* are the height and width of the beams, respectively, and *d* is the width of the gap. $p_r$ is the pressure in the resist, $p_i$ is the pressure at the air-resist interface, and the air pressure $p_0$ is constant during resist filling process. *θ* is the contact angle between the mold and the resist. (c) Schematic of the calculation of maximum capillary force induced stress across a beam of the mold. The maximum stress across the beam is located at the corners. (d) Schematic of model structures in the FEM simulation. The simulation consisted of two steps, a two-phase flow (TPF) model, and a solid mechanics model. In the TPF model, the *x*- and *y*-direction of the resist are the inlets, and the top of the channel is the outlet. The walls are wetting walls with contact angles of 30°. In the solid mechanics model, the mold is considered as an elastic material, and the pressure difference obtained in the previous step is applied to the walls of beams.

## 3. Results and discussion

There are several independent variables that must be specified to solve in our model. These variables include the contact angle of the resist liquid with the mold *θ*, the surface tension of resist liquid *γ*, the thickness of resist film *H*, Young's modulus *E* and yield strength $\sigma_Y$ of mold material, and two structure dimensions w, h and d. The numerical values of these physical parameters used are listed in **Table 1**.

Table 1 Numerical values of the physical parameters used in FEM simulation and theoretical calculation.

| Physical parameter | Symbol | Value |
| --- | --- | --- |
| Surface tension (N·m$^{-1}$) | *γ* | 0.03 |
| Resist viscosity (Pa·s) | *μ* | 0.01 |
| Resist thickness (nm) | *T* | 400 |
| Ambient pressure (bar) | $p_0$ | 1.01 |
| Pressure inside the resist | $p_r$ | |
| Pressure at the resist-air interface | $p_i$ | |
| Contact angle | *θ* | 45°, 30° |
| Density of resist (kg·m$^{-3}$) | *ρ* | $1.0 \times 10^3$ |

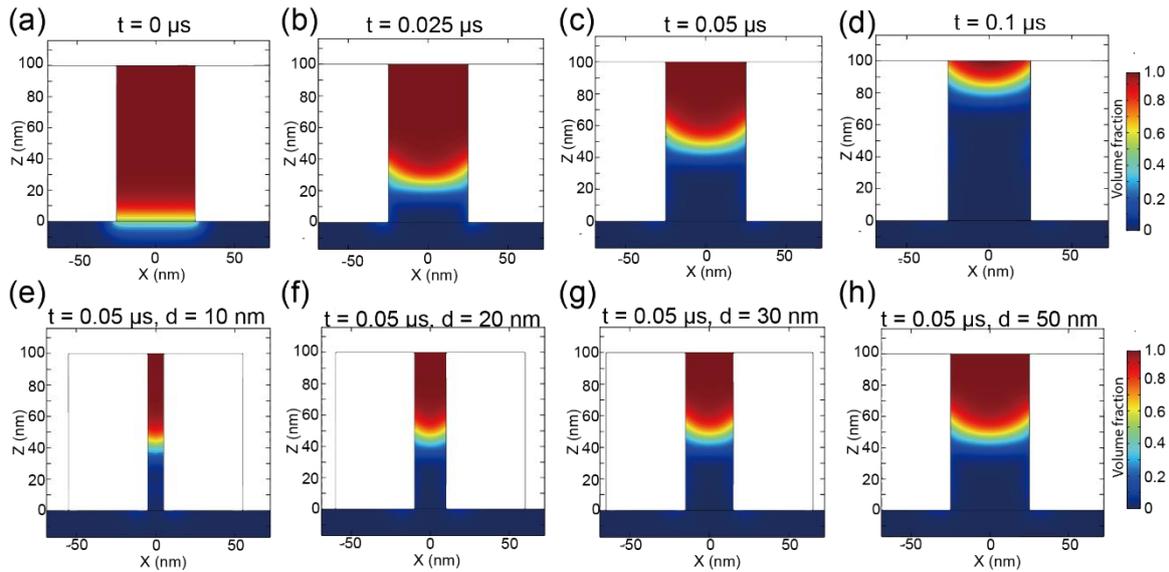

Figure 2 (a) Simulated volume fraction of the resist flow in a 100-nm-height channel with a 50-nm-width gap between two beams on the mold during the resist filling process at a different time. (b) The simulated volume fraction of the resist flow in 100-nm-height channels with the gap width range from 10 nm to 50 nm at time t = 0.05 µs.

*3.1 Tim-dependent imprint resist filling process*

The dynamic filling process of imprint resist induced by the capillary force leads to a pressure difference across the beam. The filling process of imprint resist can be simulated by TPF in FEM and the time-dependent filling process of the resist flow in a 50 nm width, 100 nm height channel of different times are shown in **Figure 2a-d**, respectively. The results clearly suggested that the resist filling process during the UV-NIL is relatively fast (< 0.5 µs), which is also consistent with the theoretical model (**Equation 5**). The filling process takes only 0.1 µs to achieve ~90% filling ratio. The impact of the gap width on the resist filling behavior was further investigated by changing the gap width in FEM simulations. **Figure 2e-f** display the simulated volume fraction of the resist flow in 100 nm height channels with different gap widths ranging from 10 nm to 50 nm at *t = 0.05 µs*. The filling ratio for narrower channel is lower than the wider channel at the same filling time because of the higher viscous resistance. However, the differences in front-tracking method-based FEM simulations are smaller than the results derived from the Lucas-Washburn equation, which is attributed to the head loss during the filling of resist.

*3.2 Tim-dependent stress distribution in the mold*

The pressure differences across the beams apply unbalanced load on the beams. To numerically analyze the elastic deformation of the beams, FEM simulations were conducted using the data obtained from hydrodynamic simulations. The von Mises distribution in the 50-nm-width beams due to the resist flow in a 50 nm width channel at t = 0.05 µs is plotted in **Figure 3a**. The maximum stress occurs at the corners, which is in correspondence with previous discussions. The stress increases with the filling ratio, due to the increase of the interaction area between the resist flow and mold beams. The relationship between the filling time and the maximum von Mises stress across the beams with different gap widths is summarized in **Figure 3b**. Apparently, the maximum von Mises stress increases with the filling ratio and reaches the maximum value at ~ 100 % filling ratio, which is also consistent with the theoretical analysis. The relationship of the maximum von Mises stress versus the gap width at ~ 100 % filling is then summarized in **Figure 3c**. We fitted the FEM simulated data using an inverse proportion model, and the results clearly show that the maximum stress is inversely proportional to the gap width, as predicted in **Equation (9)**.

*3.3 Comparison between the FEM simulation and theoretical calculation*

The theoretical analysis according to **Equation (9) and (10)** implies that for typical grating structures, the maximum stress distributed across the beam on the mold reaches its maximum value at nearly 100% filling, and the stress distribution can be obtained from FEM simulations. However, FEM simulation is time-consuming hence not suitable for practical evaluation. Therefore, in addition to the FEM simulation, we have proposed a theoretical model for quick assessment of the capillary force induced stress in the mold structures in nanoscale UV-NIL, based on **Equation (10)** and compared with the FEM simulations for cross-validation of the model. **Figure 4** shows the maximum von Mises stress in the beams as a function of the height of the beams for 5 nm (**Figure 4a**) and 100 nm (**Figure 4b**) width parallel beams obtained from FEM simulations and theoretical calculations, respectively. The results show that for both cases, the model agrees well with the FEM simulation, which proves that our model is suitable for the evaluation of capillary force induced stress distribution in mold structures in UV-NIL from sub-10 nm to 100 nm resolutions.

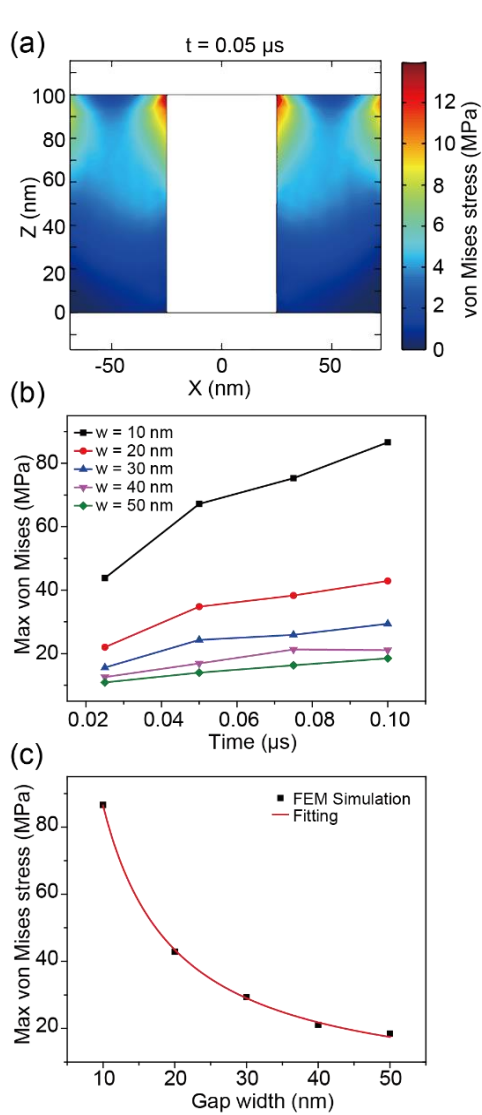

Figure 4 Time-dependent stress distribution inside the mold. (a) Von Mises stress distribution inside two 50-nm-width, 100-nm-height beams induced by the resist flow in the channel between the beams at t = 0.05 μs. (b) Max von Mises stress across the beam at different filling stages with beam widths ranging from 10 nm to 50 nm. The heights of the beams were held at 100 nm. (c) Max von Mises stress inside the beam for different gap widths near 100% filling ratio. The widths and heights of the beam were held at 50 nm and 100 nm. The data is fitted using an inverse proportion model.

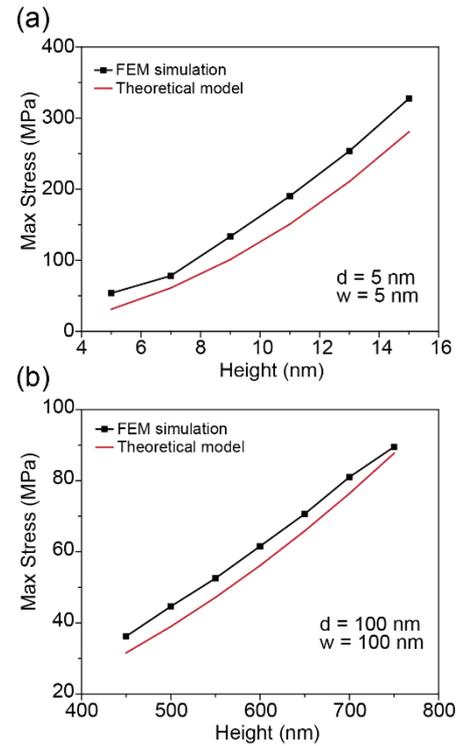

Figure 3 Plot of the maximum von Mises stress in two parallel beams as a function of the height of the beams for (a) 5 nm and (b) 100 nm width beams obtained from FEM simulation (black squares) and theoretical calculation (red circles), respectively. The heights of the beams were held at 100 nm.

*3.4 Effects of physical parameters of resist and mold materials*

In principle, a narrower beam can lead to a smaller momentum of inertia, and a higher beam results in a larger moment, which both cause a higher level of stress distribution. **Figure 5a** and **5b** plots the maximum stress as a function of gap width and width of gratings, respectively. The results evidence that the gap width and height of mold beams are the dominant factors for the stress distribution during the UV-NIL process.

Physical parameters of resist liquid also play an essential role in the imprint process. As discussed in the previous paragraphs, the viscosity of the resist flow affects only the filling process but not the stress distribution in the mold structures. However, the surface tension and contact angle of resist liquid change the maximum stress inside the beams significantly. The relationships of surface tension and contact angle with the maximum stress inside two 100 nm width, 100 nm gap width beams are summerized in **Figure 5c** and **5d**, respectively. The results indicate that the maximum stress increases linearly with the surface tension, due to the increased capillary force and Laplace pressure, and decreases with the contact angle, because lower contact angle leads to better wettability and increases the capillary force accordingly.

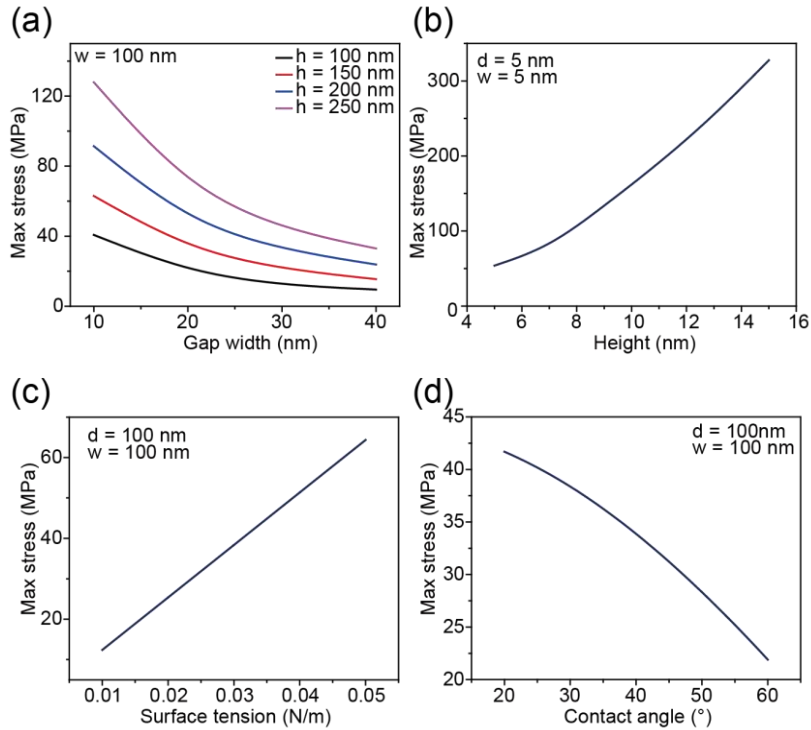

Figure 5 Effects of physical parameters of the resist and mold materials on the maximum stress inside the mold structures, showing the maximum stress versus (a) gap width between the beams, (b) height of the beams, (c) surface tension of the resist, and (d) contact angle between the resist and the mold. The heights of the beams were fixed at 200 nm in (c) and (d).

*3.5 Critical aspect ratio of the imprint mold*

According solid mechanics theory, structures collapse when the maximum stress exceeds the yield strength of materials. As illustrated in **Figure 1c**, the maximum stress occurs at the corner of a beam of the imprint mold during the imprinting process. The stress distributed at the corner near the resist liquid flow is compressive; while the stress at the opposite corner is tensile. Therefore, the mold structure collapses through the unbalanced loading. The collapsing process could be predicted by comparing the maximum stress with the yield strength of the mold material. For different mold materials, the yield strength varies significantly, e.g., 7000 MPa for silicon, 364 MPa for silicon dioxide. For the mold used in sub-10 nm UV-NIL made of crosslinked HSQ using HIBL, we assume the yield strength of HSQ ranges from 60 – 100 MPa, taking the adhesion of HSQ structures with the substrate into consideration.

**Figure 6a** shows the critical aspect ratio as a function of gap width of a two-parallel-beams imprint mold for different yield strength ranging from 60 MPa to 100 MPa, the result indicates that the highest aspect ratio allowed for transferring the 10 nm period beams to the resist through UV-NIL is only 1.8, even for a mold with yield strength of $\sigma_Y = 100$ MPa. Therefore, to improve the critical aspect ratio, materials with higher yield strength and adhesive strength is required.

For non-wetting UV-NIL, an imprint pressure is required to overcome the capillary force at the resist-wall interface. The critical pressure can be estimated according to **Equation (12)**. **Figure 6b** shows the critical imprint pressure as a function of the gap width of the beams on the non-wetting imprint mold in UV-NIL, and the width of the beams is the same as the gap width. The critical pressure is calculated for a contact angle of 110°, and the result indicates that critical imprint pressure increases when the gap width of the beams shrinks. For UV-NIL system with sub-10 nm structure, a critical imprint pressure > 4 MPa is required. The relationship of the critical aspect ratio for non-wetting molds versus the gap width of the beams is plotted in **Figure 6c**. The critical aspect ratio increases with yield strength but decreases with contact angle of the resist with the mold.

**4. Conclusion**

In conclusion, we have established a model to describe the dynamic mechanical behavior in UV-NIL towards sub-10 nm resolution. We have theoretically and numerically investigated the capillary force induced fluidic flow of resist and pattern deformation in UV-NIL. The maximum stress and critical aspect ratio in both wetting and non-wetting molds

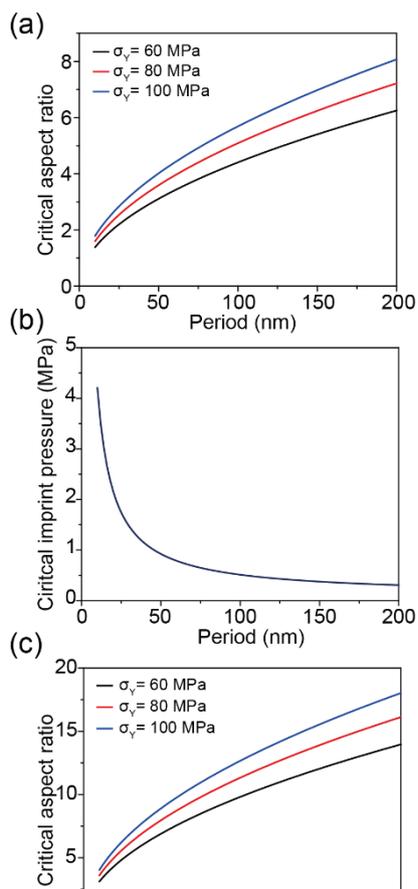

Figure 6 (a) Critical aspect ratio as a function of the period for the wetting imprint mold with different yield strengths of 60 MPa, 80 MPa, and 100 MPa, respectively. (b) Critical imprint pressure as a function of the period for the non-wetting imprint mold. (c) Critical aspect ratio of the beams as a function of the period of the non-wetting imprint mold with different yield strengths of 60 MPa, 80 MPa, and 100 MPa, respectively.

have been calculated for pattern structures ranging from 5 nm to 200 nm. This study has reported here the key factors affecting the durability of imprint molds and provided guidelines for reducing the pattern deformation on the imprint mold. Based on our study, the critical aspect ratio for grating patterns with a given period is limited by both the mechanical properties of the resist and the mold. The presented model will provide a guideline for proper selection of resist and optimization of the aspect ratio of the nanostructures on the mold to achieve high-fidelity replication of nanostructures at sub-10 nm resolution.

## Acknowledgements